\documentclass[sigconf]{acmart}

\usepackage{booktabs} 

\usepackage[margins]{trackchanges}

\usepackage[colorinlistoftodos]{todonotes}
\setcopyright{rightsretained}

\addeditor{MF}
\addeditor{EE}

\begin{document}
\title{Test Agents: Adaptive, Autonomous and Intelligent Test Cases}

\author{Eduard Paul Enoiu and Mirgita Frasheri}
 \affiliation{%
   \institution{M\"alardalen University, V\"aster{\aa}s,  Sweden.}
 }


\begin{abstract}
Growth of software size, lack of resources to perform regression testing, and failure to detect bugs faster have seen increased reliance on continuous integration and test automation. Even with greater hardware and software resources dedicated to test automation, software testing is faced with enormous challenges, resulting in increased dependence on complex mechanisms for automated test case selection and prioritization as part of a continuous integration framework. These mechanisms are currently using simple entities called test cases that are concretely realized as executable scripts. Our key idea is to provide test cases with more reasoning, adaptive behavior and learning capabilities by using the concepts of intelligent software agents. We refer to such test cases as {\it test agents}. The model that underlie a test agent is capable of flexible and autonomous actions in order to meet overall testing objectives. Our goal is to increase the decentralization of regression testing by letting test agents to know for themselves when they should be executing, how they should update their purpose, and when they should interact with each other. In this paper, we envision software test agents that display such adaptive autonomous behavior. Emerging developments and challenges regarding the use of test agents are explored---in particular, new research that seeks to use adaptive autonomous agents in software testing.
\end{abstract}

%
%


\maketitle

\section{Introduction}
Even if software testing is widely used in industry for verification and validation, in many cases due to the increased use of continuous integration and the sheer amount of test cases created, automation becomes a bottleneck in software development and is expensive to perform in a cost-efficient manner. Several such challenges have been identified in the automated regression testing of complex software systems \cite{strandberg2017automated,memon2017taming}: costly scheduling of test cases, badly prioritized test suite, and forgotten test cases. Automated testing is the process of designing, continuously executing and maintaining the confidence in the system dependability in a cost-effective and automated manner. In this context, test cases are created by human testers satisfying different test requirements and domain needs, are scripted and executed automatically and repeatedly. These test cases contain some mechanism for test evaluation that is embedded in a test script. 
Traditional regression test selection mechanisms are not designed to exhibit capabilities of responsiveness, flexibility, robustness and re-reconfigurability, since they are built upon centralized systems that strive to achieve overall test suite optimization, but have a weak and rigid response to complexity and changes at runtime. Such centralized regression testing mechanisms normally lead to situations where test cases are not adapting, resulting in inefficient and costly test scheduling mechanisms. In these circumstances, the current challenge is to develop collaborative and reconfigurable test cases that support characteristics of adaptation, autonomy and intelligence. 

In this paper we outline our vision \textit{to decentralize test automation and the control of regression testing by developing tests cases that are capable of intelligent, autonomous and adaptive actions}. Such test cases are named \textit{test agents}. We envision the use of a test agent as a self-contained and self-aware test case capable of interactions with other test agents. These test agents represent another way for an engineer to design test cases that will effectively test software. The use of test agents tackles these test automation challenges by enabling test engineers to create autonomous and adaptive test cases which can take decisions about their action-execution mechanism and scope of interactions at run-time.

Many possible definitions of agents exist in the literature \cite{franklin1996agent}. Here we explicitly consider that agents are software systems that operate in an execution environment which they can perceive and respond to, take initiatives and select own goals, and interact with others when deemed fit \cite{jennings1998applications}. 
Over the past few years, this paradigm has been applied in different application domains, with varying levels of product maturity \cite{muller2014application}. These solutions are being deployed in the telecommunication, logistics, e-commerce, and robotics domain. Having learned from the successes and drawbacks of using agents in other domains, our vision is to introduce and implement the test agent paradigm. Practically, our vision goals are to: (i) help test engineers create test cases capable of adaptive and autonomous actions using test agents, (ii) develop a language for describing test agents, i.e., agents with a specific purpose in terms of test effectiveness and efficiency, a set of interaction and execution mechanisms, and the ability to perceive the test evaluation results after each run, and (iii) investigate how test agents and their interactions evolving in time could be represented using specific rules.

\section{Regression Test Automation}
\begin{figure}
   \centering
     \includegraphics[width=0.47\textwidth]{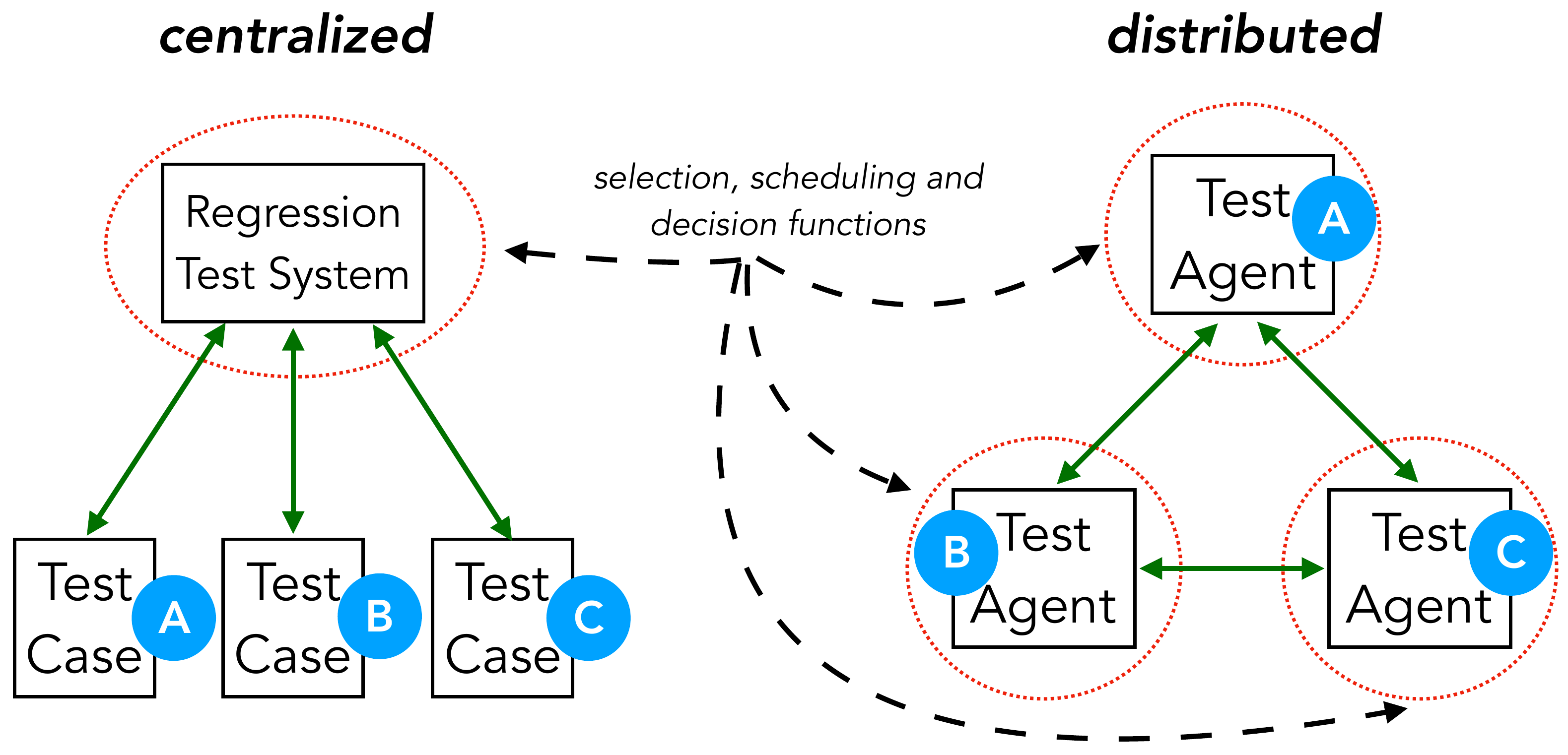}
   \caption{Centralized and distributed approaches to regression test automation.}\label{overall}
 \end{figure}

Software testing is the primary method used in industrial practice to evaluate software and can be divided \cite{ammann2016introduction} in three distinct tasks: test design, test execution and test evaluation. A test engineer designs tests by creating test requirements which are then written into actual scripts that are ready for execution. These scripts are executed against the software and the results are evaluated. Test automation is using software to control these activities with the aim to reduce the cost of testing. One integral part of test automation is regression testing, the process of continuously testing software that has been modified. 
A regression test system (shown in Figure \ref{overall}) is often incorporated into a continuous integration development and determines which test cases to include in a regression suite by identifying suitable cases based on different information sources (i.e., fault history, execution time, test coverage, failing tests) obtained after the execution of the system. In the current practice of software testing, test cases are entities composed of several discrete parts (i.e., test case input values and expected results needed for evaluation). These components are concretely realized in a script that can be automatically executed and knows exactly what values to expect. As a result, the existing process of software testing is build upon static and rather simplistic test cases, thus entailing the use of a highly-complex and centralized test scheduling technique for regression testing. To change this centralized process, we envision a new class of autonomous, intelligent and adaptive test cases that we refer to as test agents. As a result, test agents could enable testing of goals beyond their original scope and can decide what interactions are needed with other test agents and adapt when their test goal or the software updates. The change from the traditional centralized approach to regression testing to the new distributed, adaptive and intelligent approach is illustrated in Figure \ref{overall}.

\section{Adaptive Autonomy}
The notion of adaptive autonomy refers to the ability of software agents to change their levels of autonomy based on their circumstances. Agent autonomy in itself can be described through two dimensions, self-sufficiency, i.e. ability to fulfill a task without outside help, and self-directness, i.e. ability to decide upon one's own goals \cite{johnson2011fundamental}. Castelfranchi \cite{castelfranchi2000founding} uses dependence theory to define autonomy as follows: An agent \textit{A} that lacks means for performing a specific task \textit{T} and depends on an agent \textit{B} to acquire such means, is said to be non-autonomous from \textit{B} with respect to \textit{T}. It might happen that \textit{A} is able to perform \textit{T} by itself at a point in time $t_1$, but not at $t_2$ due to circumstantial changes, e.g. \textit{A} is low on resource-consumption levels. Consequently, \textit{A} (and \textit{B} as well) needs to continuously evaluate whether it needs assistance, or whether it is willing to give assistance to other agents that might ask. As a result, based on their circumstances, agents decide by themselves when to adapt their autonomy. Alongside adaptive autonomy, there are other similar notions such as adjustable autonomy \cite{hardin2009using} \cite{johnson2011fundamental}, mixed-initiative interaction \cite{hardin2009using}, collaborative control \cite{fong2001collaborative} and sliding autonomy \cite{brookshire2004preliminary}. 

\section{Agents in Software Testing}
Software agents have already been used to automate different aspects of testing. One such approach is the adaptive test management system (ATMS) \cite{malz2010agent}, which aims at selecting an appropriate set of test cases to be executed in every test cycle. The ATMS uses three types of agents: test unit, test case and test resource. Test unit agents request additional test cases when the unit does not fulfill the desired test coverage using fuzzy logic. Since changes made to a software unit could influence other units, agents exchange information with each other about such events. Agents calculate the local priority of their respective test cases, and negotiate with each other on global priority using an action selection scheme. 
Researchers have also used a multi-agent approach for intra-class
testing of object-oriented software. Dhavachelvan \cite{dhavachelvan2006new,dhavachelvan2005multi} presented three types of agents: distributor agent, testing technique agent, and clones. Distributor agents take assignments and map them to the available testing agents. Agents are able to clone themselves to accommodate the resource needs required by specific testing activities. In this approach, testing agents do not communicate with each other, but only with their distributors. 
Other contributions have relied on the agent-based paradigm to specifically target service-oriented systems \cite{bai2011design}. The Belief Desire Intention (BDI) agent architecture is used by Rao \textit{et al.} \cite{rao1995bdi} to distinguish between two types of agents: coordinators and runners. Coordinators create testing plans and runners conduct the testing activities and send their results back to a coordinator. 
Hong Zhu \cite{zhu2004cooperative} is using the agent-paradigm in a framework that targets both software development and management. Zhang \textit{et al.} \cite{zhang2008mobile} extended the LoadRunner testing platform for web services using IBM Aglet agents. LoadRunner enables the simulation of users by executing tests on the remote server hosting the service by using Aglets agents.
A different approach is presented in the work of Tang \textit{et al.} \cite{tang2010towards}. Their study aims at automating the whole testing life cycle by using four types of agents: requirement agent (i.e., a mapping between software and test requirements), construct agent (i.e., generation of test cases) and an execution and report agent.

To summarize, what is missing from the state of the art is a wide-ranging approach for test automation in the context of regression testing, such that test cases carry out actions with some degree of autonomy and adaptivity. Test agents will tackle this gap and expand the scope of application of software agents to regression testing and test automation.

\section{Defining a Test Agent}
Our overall vision for the use of \textit{test agents} is \textit{to shift the bulk of continuous test selection, prioritisation and scheduling from a centralized regression test automation framework to a lower level of abstraction where test agents can decide by themselves how and what to execute}. 

\subsection{A Test Agent Model}
\begin{figure}
  \centering
    \includegraphics[width=0.35\textwidth]{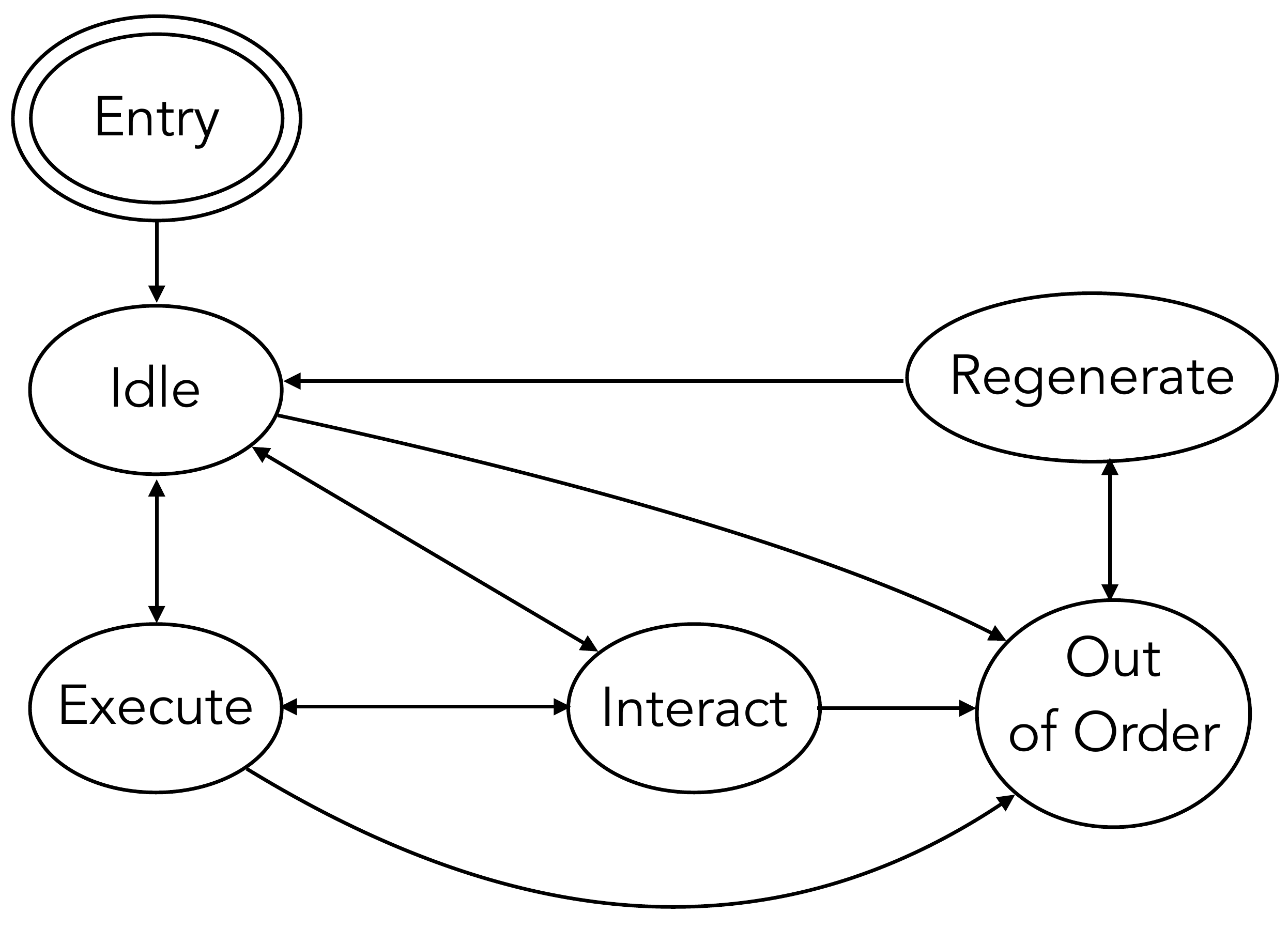}
  \caption{A high-level test agent model.}\label{fsm}
\end{figure}

A test agent model can be realized by considering the problem on two levels of abstraction. The high-level is concerned with modeling the general internal operation of the test agent, whereas the low-level (i.e., behavioral level) tackles the modeling of mechanisms that allow the test agent to adapt its autonomy, thus displaying adaptive autonomous behavior. The test agent is composed of the following five states mirrored in Figure \ref{fsm}: \textit{Idle}, \textit{Interact}, \textit{Execute}, \textit{Regenerate}, and \textit{Out of Order}. The test agent is not committed to anything in the \textit{Idle} state and will execute its own task when needed. Once the execution task has been generated, or the test agent has gotten a task from another test agent, it will go to the \textit{Execute} state and decide if it needs assistance from another test agent before and after its execution. 
After the execution is completed, the agent will go back to the \textit{Idle} state. When the test agent receives a request from another agent, it will switch to the \textit{Interact} state and will decide whether to accept the request and give assistance or discard it. When the test agent decides it cannot serve its initial purpose it will switch to the \textit{Out of Order} state. Other triggers for switching to \textit{Out of Order} could be devised if necessary. In addition, the agent can switch to the \textit{Regenerate} state and a test case redesign takes place with the help of a test engineer. In the end, the test agent can return to the \textit{Idle} state.  For example, let us assume a test agent A has not been able to fulfill its original goal (e.g., achieving 100\% branch coverage for a certain function) due to a code change. The agent will ask for assistance at runtime in the \textit{Execute} state from another test agent. Test agent C receives the request, but decides to discard the request since its initial goal was to check the fulfillment of a certain requirement and its last execution is not affecting the logic that needs to be covered. Test agent B decides to accept the request since this new goal serves its initial purpose and goes to the \textit{Execute} state and fulfills it. 

\subsection{Test Agent Interactions}
The adaptive autonomous behavior is determined at those points in which the test agent decides whether to ask or give help, and is modeled through its willingness to interact. This willingness is composed of the disposition to give or to ask for help. Frasheri \textit{et al.} \cite{frasherijournal} considered the different factors that could influence the willingness of agents to interact, while Van der Vecht \textit{et al.} \cite{van2009autonomy} examined the task urgency and agent dedication to the overall organization as a molder of adaptive behavior. Further studies are crucial for establishing a suitable adaptive autonomous behavior based on the application of test agents in realistic testing contexts, where agents cannot be expected to accurately interact.

In this paper, we propose to derive the adaptive behavior of test agents by adapting the following four levels of interactions between agents already identified by Frasheri \textit{et al.} \cite{frasheri2017towards} to distributed regression testing: (i) non-committal interactions in which a test agent can broadcast information (e.g, its execution time, fault detection, test coverage) to the other test agents and no response is expected, (ii) one-to-one dialogue in which a test agent \textit{A} asks another test agent \textit{B} for information (e.g., its fault history) and a response is expected, (iii) one-to-one delegation which is used when a test agent \textit{A} delegates a task, or a subtask (e.g., cover certain parts of the code) to a test agent \textit{B} and a response is expected together with some execution evidence and information (e.g., the input parameters used during execution), and (iv) one-to-many dialogue/delegation in which two scenarios are considered: chain interactions and simultaneous interactions (in the former, a test agent \textit{A} makes a request to a test agent \textit{B}, which in turn makes a request to \textit{C}; and whereas in the latter, test agent \textit{A} makes several requests, one to test agent \textit{B}, one to \textit{C}). For example, the one-to-many interactions can be used to achieve a trade-off between multiple test agents and their objectives with regard to some test criteria and cost (e.g., maximize test coverage, minimize the execution time).

\section{Challenges}
Reassessing the concept of a test case using a test agent representation is not an easy task to accomplish and therefore realizing our vision for the use of test agents in regression testing requires addressing the following challenges.


\subsection*{Design of Test Agents}
When it comes to creating test cases there are at least two ways \cite{ammann2016introduction}: criteria-based and human-based test design. The criteria-based test design is used for creating tests that satisfy some test requirement or coverage criterion. This process requires the creation of explicit test requirements and models. On the other hand, human-based test design is used for creating test cases based on the test engineer's domain-specific knowledge. When engineers create tests, they sometimes attempt to perform positive testing as well as stressing the software using unusual test cases. One challenge to this end is to provide precise guidance to test engineers on how to create a test agent in terms of its purpose, test case values, execution environment, perception capabilities and interaction actions with other test agents. The design of test agents is complicated by the test case heterogeneity given the large space of possible test scenarios and interactions with the software. Developing a programming language for expressing test agents depends on the software under test, the nature of the test design techniques used and the types of faults targeted by testing. A challenge is therefore to define interaction rules for test agents as well as a language to describe its perception capabilities, interaction rules for actions, test purpose and test agent hierarchy. 

\subsection*{Test Automation}
We refer to a test being automated if its execution, evaluation and reporting is controlled by software. As an example, when dealing with test agents, test automation necessarily has to consider a standardized design for test scripts, and should include support for a test execution driver. This driver should be used by each test agent for executing the software, evaluate the results of its execution and report the results back to the test agent. A challenge is to establish a test automation framework that supports (i) the ability to share test data and interaction information among test agents, (ii) the ability for test agents to easily organize and run, and (iii) statistical assertions to evaluate the multi-dimensional information perceived from logs and reports. Clearly, automated support for maintaining test agents is crucial for the success of such an approach. 

\subsection*{Regression Test Agent Selection}
Software is subject to frequent modifications. Regression testing is the process of continuously testing modified software. Its purpose is to ensure that software is functionally equivalent to the version before the updates. For example, regression testing can reveal if mistakes in requirements are implemented in the software. The use of regression testing can result in a test suite that is too large to manage and does not finish to execute in a timely manner. 
For test agents, regression testing is associated with the interaction between agents and their evolution in time. Evolving a test agent is challenging because of more complex dependencies. The adaptive autonomous behavior of test agents is modeled through its willingness to interact with other test agents. This interaction should be based on local built-in preferences that are deciding what to do next and initiate actions during runtime.

\section{Overall Objectives and Conclusion}
The goal of this work is to apply the adaptive autonomous agent paradigm to the software testing domain in order to reassess the notion of a test case. A test agent is more intelligent than a test case because it behaves as a dynamic entity that can decide by itself or in a group of agents how and what the software should execute at runtime. The vision proposed in this paper is expected to lead to an operational definition of a test agent. Such agents need to continuously reason and decide on their need for helping or assisting other test agents in different circumstances. 

In other words, the vision is for test engineers to create test agents that carry out a set of actions with some degree of autonomy by interacting with other agents and driven by the hard-coded knowledge of a test engineer's goals. In order to validate the proposed vision, the following steps need to be taken: (i) select a platform in which to develop the test agent automation system (i.e., using several agent-based technologies are available such as JADE, NetLogo, SeSAm \cite{kravari2015survey}), (ii) analyze and simulate (e.g., using ROS (Robot Operating System) \cite{quigley2009ros}) how test agent interactions are shaped by a test engineer's preferences and define how interactions between different agents are represented, and (iii)
investigate different learning techniques that can help the test agent refine its decision-making process and evolution in time. 

An advanced capability that can be added to test agents is learning such that they retain useful information from their interactions as training data and utilize various machine learning techniques to adapt to new execution scenarios and improve their performance.


\bibliographystyle{ACM-Reference-Format}
\bibliography{acmart} 

\end{document}